\documentclass[twocolumn,floats,showpacs,superscriptaddress,pre]{revtex4}
\usepackage{amsmath,amssymb,bm}
\usepackage{graphics}
\usepackage{epsfig}
\usepackage{graphicx}

\begin{document}

\title{Quantum Action Principle for Covariant Systems. Bosonic string.}
\author{Natalia Gorobey and Alexander Lukyanenko}
\email{alex.lukyan@rambler.ru}
\affiliation{Department of Experimental Physics, St. Petersburg State Polytechnical
University, Polytekhnicheskaya 29, 195251, St. Petersburg, Russia}

\begin{abstract}
A new form of covariant quantum theory based on a quantum version of the
action principle is considered for the case of a free bosonic string. The
central idea of the new approach is to delay conditions of stationarity of
the classical action with respect to Lagrangian multipliers up to the
quantum level where delayed conditions of stationarity are imposed on a
quantum action. Physical states of the ordinary covariant quantum theory are
replaced by well defined stationary states as those which obey quantum action
principle. The stationary states have well defined energies.
\end{abstract}

\maketitle
\date{\today }





\section{\textbf{INTRODUCTION}}

The present work is devoted to the development of a new approach to
covariant quantum theory based on a quantum version of the action principle.
A wide class of covariant theories includes gauge theories and precisely
covariant theories with quadratic constraints on canonical momenta such as
General Relativity and string theory. General canonical analysis and a
variant of quantization of covariant theories was proposed by Dirac \cite{D}%
. Classical action of a covariant theory has the canonical form:
\begin{equation}
I=\int d\tau \left( p_{\alpha }\overset{\cdot }{q}_{\alpha }-\lambda
_{a}C_{a}\right) ,  \label{1}
\end{equation}%
where $C_{a}$ are constraints, and $\lambda _{a}$ are Lagrangian
multipliers. Conditions of stationarity of the classical action (\ref{1})
with respect to Lagrangian multipliers $\lambda _{a}$ give the classical
constraint equations,
\begin{equation}
C_{a}=0.  \label{2}
\end{equation}
In Dirac approach to covariant quantum theory, which may be called
"ordinary" covariant approach, canonical variables $q_{\alpha }$ and $p_{\alpha }$
are replaced as usual by operators, but Schr\"{o}dinger wave equation is
replaced by the system of quantum constraints:
\begin{equation}
\widehat{C}_{a}\psi =0.  \label{3}
\end{equation}%
The set of wave equations (\ref{3}) ensures the covariance at the quantum
level. A solution $\psi \left( q\right) $ of this set, being independent
on a choice of Lagrangian multipliers $\lambda _{a}$, corresponds to a
"physical state". It is not our goal to discuss all problems of the ordinary
covariant quantum theory. The main problem is the problem of consistency of
the system. The second problem is the problem of dynamical and
probabilistic interpretation of physical states. Even in the simplest
covariant quantum theory, i.e., one-particle relativistic quantum mechanics
(RQM), we meet these problems \cite{BD}.

A new approach to quantization of the dynamics of relativistic particle,
which solves the problem of probabilistic interpretation of RQM, was
proposed in \cite{GLL1,GLL2}. We call this new approach as a ''quantum action
principle (QAP)´´. In a general case, the main idea of QAP is to "delay"
the conditions of the stationarity of the classical action with respect to
Lagrangian multipliers $\lambda _{a}$ up to the quantum level. At the
quantum level in place of the system (\ref{3}) we write the ordinary Schr%
\"{o}dinger equation in the internal time parameter $\tau $:
\begin{equation}
i\hbar \frac{\partial \psi }{\partial \tau }=\lambda _{a}\widehat{C}_{a}\psi
.  \label{4}
\end{equation}%
Now the problem of the consistency of this set of equations is absent. A
solution of the equation (\ref{4}) defines an amplitude $K\equiv
\left\langle out\right\vert \left. in\right\rangle $ for some quantum
transition (for example, in real experiment) $\left\vert in\right\rangle
\rightarrow \left\vert out\right\rangle $. Writing the transition amplitude
in the exponential form,
\begin{equation}
K=\exp \left( \frac{i}{\hbar }\Lambda +R\right) ,  \label{5}
\end{equation}%
we consider its real phase function $\Lambda $ as a quantum action,
corresponding to the quantum transition $\left\vert in\right\rangle
\rightarrow \left\vert out\right\rangle $. Being a functional of the Lagrangian
multipliers $\lambda _{a}$ (and kinematical parameters of the experiment),
the quantum action $\Lambda $ gives us a possibility to fix the Lagrangian
multipliers $\lambda _{a}$ by means of the "delayed" conditions of the stationarity:
\begin{equation}
\frac{\delta \Lambda }{\delta \lambda _{a}}=0.  \label{6}
\end{equation}%
From Eq. (\ref{6}) we obtain  Lagrangian multipliers $\lambda _{a}$ (more
precisely, some invariants formed by $\lambda _{a}$) as functions of
kinematical parameters of the experiment. The only memory of quantum
anomalies in our approach is the question: to what extent the covariance is
conserved by the delayed conditions of the stationarity (\ref{6})? In any case, the
transition amplitude (\ref{5}) is well defined. Let us stress, that the
equation (\ref{4}) describes the dynamics of the system, in so far as
kinematical parameters of boundary states $\left\vert in\right\rangle$ and
$\left\vert out\right\rangle $ include the $x^{0}$- coordinate of the
Minkowsky space.

In the present work we develop the variant of QAP proposed in \cite{GLL2} 
in application to quantum theory of a closed bosonic string. In this
approach, QAP is supplemented by a set of additional conditions which make
the dynamics of the $x^{0}$- coordinate to be classical. This modification of QAP
does not destroy the covariance of the theory, but it restores the explicit
dynamical contents of the theory and solves the problem of
probabilistic interpretation. In place of physical states defined
by the set of equations (\ref{3}) (if they exist) we propose a definition of
stationary states as states which satisfy the principle of the stationarity of the
quantum action $\Lambda $. In contrast to the ordinary definition of the
spectrum of excitations of a string in fixed gauges (the light-cone gauge is
the most wide used), our definition of stationary states is formally
covariant. In the stationary states, the energy of a string takes certain
values.

In the next section we will apply QAP to the well known simplest covariant
object, i.e., a relativistic particle. Latter the new approach will be applied
to a free bosonic string.

\section{QUANTUM ACTION PRINCIPLE IN RELATIVISTIC QUANTUM MECHANICS}

We begin with the classical action of a relativistic particle in the
canonical form
\begin{equation}
I=\int\limits_{0}^{1}d\tau \left( p_{\mu }\overset{\cdot }{x}^{\mu
}-NH\right) ,  \label{7}
\end{equation}%
where
\begin{equation}
H\equiv p_{\mu }p^{\mu }-m^{2}c^{2}\approx 0  \label{8}
\end{equation}%
is the Hamiltonian constraint equation, which is the Euler-Lagrange (EL)
equation for the Lagrangian multiplier $N$. The action (\ref{7}) is
the reparameterization invariant. The constraint equation (\ref{8}) is a
restriction on the initial data in a phase space of a particle, and it does not
define $N$. The "wavy" equality means, that the equation (\ref{8}) has to be
solved at the final stage, when all dynamical equations are taken into
account. In fact, this means calculation of a stationary value of the
classical action which may be performed in its Lagrangian form. If we take
into account the EL equation for the momentum variable $p_{\mu }$,
\begin{equation}
\overset{\cdot }{x}^{\mu }-2Np^{\mu }=0,  \label{9}
\end{equation}%
we return to the Lagrangian form of the classical action
\begin{equation}
I=\int\limits_{0}^{1}d\tau \left( \frac{\overset{\cdot }{x}^{2}}{4N}%
+m^{2}c^{2}N\right) .  \label{10}
\end{equation}%
The action (\ref{10}) also is a reparameterization invariant, but now
the condition of its stationarity with respect to the variable $N$ fixes the
reparameterization invariant integral:
\begin{equation}
T\equiv \int\limits_{0}^{1}d\tau N=\frac{\sqrt{\left( x_{1}-x_{0}\right)
^{2}}}{2mc},  \label{11}
\end{equation}%
where $x_{0,1}^{\mu }$ are end points of a world line of a particle.
In this sense the Lagrangian form of the classical action is more
fundamental, than canonical one \cite{D1}. The departure of the stationary
conditions in two forms of the classical action will be traced in QAP which
gives us a quantum analog of the Lagrangian form of the classical action.
However, quantum mechanics is based on the canonical form of classical
theory.

Let us turn to quantum mechanics. In the ordinary approach the classical
constrain equation (\ref{8}) is replaced by the Klein-Gordon (KG) equation
\begin{equation}
\left( \partial _{\mu }\partial ^{\mu }+\frac{m^{2}c^{2}}{\hbar ^{2}}\right)
\psi =0.  \label{12}
\end{equation}%
The problem of interpretation of its solutions was mentioned above.
Our proposal \cite{GLL1,GLL2}
is to delay the condition of the stationarity with respect to the Lagrangian
multiplier $N$ up to formulation of a quantum dynamical problem. Following
the logic of the work \cite{GLL2}, let us slightly modify the original
canonical action (\ref{7}) as follows:
\begin{eqnarray}
I &=&\int\limits_{0}^{1}d\tau \left[ p_{\mu }\overset{\cdot }{x}^{\mu
}-N\left( d^{2}-p_{i}^{2}-m^{2}c^{2}\right) \right.  \notag \\
&&\left. -\lambda \left( d-p_{0}\right) \right] .  \label{13}
\end{eqnarray}%
Additional variables $d\left( \tau \right)$ and $\lambda \left( \tau \right) $
might be excluded at the classical level with restoration of the original
action (\ref{7}). But we delay the restoration to the quantum level. This
enlargement of the set of variational parameters will follow the classical
character of the dynamics of the $x^{0}$-coordinate of a particle. In place
of the KG equation (\ref{12}) we write the modified Schr\"{o}dinger equation
\begin{eqnarray}
i\hbar \frac{\partial \psi }{\partial \tau } &=&\left[ Nd^{2}+\lambda \left(
d-\frac{\hbar }{i}\frac{\partial }{\partial x^{0}}\right) \right.  \notag \\
&&\left. -N\left( -\hbar ^{2}\Delta +m^{2}c^{2}\right) \right] \psi .
\label{14}
\end{eqnarray}
Let $\psi \left( \tau ,x^{\mu }\right) =\psi _{0}\left( \tau ,x^{0}\right)
\psi _{1}\left( \tau ,x^{i}\right) $. The first multiplier, which describes
quantum dynamics of the $x^{0}$-coordinate of a particle, obeys the equation
\begin{equation}
i\hbar \frac{\partial \psi _{0}}{\partial \tau }=\left[ Nd^{2}+\lambda
\left( d-\frac{\hbar }{i}\frac{\partial }{\partial x^{0}}\right) \right]
\psi _{0}.  \label{15}
\end{equation}%
We look for a solution of the equation (\ref{15}) in the exponential form:
\begin{equation}
\psi _{0}\left( \tau ,x^{0}\right) =\exp \left[ \frac{i}{\hbar }\chi \left(
\tau ,x^{0}\right) \right] ,  \label{16}
\end{equation}%
where $\chi \left( \tau ,x^{0}\right) $ is a complex phase function for
which a quadratic representation,
\begin{equation}
\chi \left( \tau ,x^{0}\right) =\chi _{0}\left( \tau \right) +\chi
_{1}\left( \tau \right) x^{0}+\frac{1}{2}\chi _{2}\left( \tau \right) \left(
x^{0}\right) ^{2},  \label{17}
\end{equation}%
is sufficient. We take $\tau =0$ at the initial moment,
\begin{equation}
\chi \left( 0,x^{0}\right) =\frac{i\hbar }{4\varepsilon ^{2}}\left(
x^{0}\right) ^{2}+p_{0}x^{0}.  \label{18}
\end{equation}%
At the final stage of calculations the limit $\varepsilon \rightarrow 0$ is
supposed. It follows from (\ref{18}), that the initial value of the $x^{0}$%
-coordinate lies in the $\varepsilon $-neighbourhood of zero, and
corresponding initial momentum equals $p_{0\text{. }}$Comparing coefficients
in both sides of the equation (\ref{14}) in front of $x^{0}$ degrees, we
obtain:
\begin{equation}
\chi _{2}\left( \tau \right) =\chi _{2}\left( 0\right) =\frac{i\hbar }{%
2\varepsilon ^{2}},  \label{19}
\end{equation}
\begin{equation}
\chi _{1}\left( \tau \right) =p_{0}+\frac{i\hbar }{2\varepsilon ^{2}}%
\int\limits_{0}^{\tau }d\widetilde{\tau }\lambda \left( \widetilde{\tau }%
\right) ,  \label{20}
\end{equation}
\begin{eqnarray}
\chi _{0}\left( \tau \right) &=&-\int\limits_{0}^{\tau }d\widetilde{\tau }%
\left[ Nd^{2}+\lambda \left( d-p_{0}\right) \right]  \notag \\
&&+\frac{i\hbar }{2\varepsilon ^{2}}\int\limits_{0}^{\tau }d\widetilde{\tau
}\int\limits_{0}^{\widetilde{\tau }}d\widetilde{\widetilde{\tau }}\lambda
\left( \widetilde{\widetilde{\tau }}\right) .  \label{21}
\end{eqnarray}
The second part of a wave function which depends on spatial coordinates
obeys the equation
\begin{equation}
i\hbar \frac{\partial \psi _{1}}{\partial \tau }=-N\left( -\hbar ^{2}\Delta
+m^{2}c^{2}\right) \psi _{1}.  \label{22}
\end{equation}%
In the present work we consider the problem of stationary states of
relativistic systems. A corresponding solution of the equation (\ref{22}) is a
plane wave with the momentum $p_{i}$:
\begin{eqnarray}
&&\psi _{1}\left( \tau ,x^{i}\right)  \label{23} \\
&=&\exp \left[ \frac{i}{\hbar }\int\limits_{0}^{\tau }d\widetilde{\tau }%
N\left( \widetilde{\tau }\right) \left( p_{i}^{2}+m^{2}c^{2}\right) -\frac{i%
}{\hbar }p_{i}x^{i}\right]  \notag
\end{eqnarray}%
Collecting together all parts of the stationary solution on the interval $%
\tau \in \left[ 0,1\right] $, we obtain real functionals in the exponent (%
\ref{5}) :
\begin{eqnarray}
\Lambda &=&-\int\limits_{0}^{1}d\widetilde{\tau }\left[ Nd^{2}+\lambda
\left( d-p_{0}\right) \right] +p_{0}\widetilde{x}^{0}  \notag \\
&&+\int\limits_{0}^{1}d\tau N\left( p_{i}^{2}+m^{2}c^{2}\right) -p_{i}x^{i},
\label{24}
\end{eqnarray}
\begin{equation}
R=-\frac{1}{4\varepsilon ^{2}}\left( \widetilde{x}^{0}+\int\limits_{0}^{1}d%
\tau \lambda \right) ^{2},  \label{25}
\end{equation}%
where $\widetilde{x}^{0}$ is a final value of the $x^{0}$-coordinate,
corresponding to the state $\left\vert out\right\rangle $. In accordance
with (\ref{25}), the dynamics of $x^{0}$-coordinate is described by a wave
packet with a width $\varepsilon \rightarrow 0$. Therefore, we have
\begin{equation}
\widetilde{x}^{0}+\int\limits_{0}^{1}d\tau \lambda =0.  \label{26}
\end{equation}%
This equation must be added to the quantum action (\ref{24}) as an
additional condition.

Now we are ready to complete the formulation of QAP, solving the problem of
the stationarity of the quantum action (\ref{24}). Corresponding stationary
equations for $\lambda $ and $d$ are
\begin{equation}
d=p_{0},2Nd+\lambda =0,  \label{27}
\end{equation}%
and, in accordance with (\ref{26}), we have a relation,
\begin{equation}
\widetilde{x}^{0}=2Tp_{0},T\equiv \int\limits_{0}^{1}d\tau N,  \label{28}
\end{equation}%
which is a quantum analog of the classical EL equation (\ref{9}). Solving (%
\ref{28}) with respect to the initial momentum $p_{0}$, and substituting this
value in (\ref{24}), we obtain:
\begin{equation}
\Lambda =\frac{\left( \widetilde{x}^{0}\right) ^{2}}{4T}+T\left(
p_{i}^{2}+m^{2}c^{2}\right) .  \label{29}
\end{equation}%
This is a quantum analog of the classical action in the Lagrangian form (\ref%
{10}). The last step is calculation of a stationary value of (\ref{29}) with
respect to $T$,
\begin{equation}
\Lambda =\widetilde{x}^{0}\sqrt{p_{i}^{2}+m^{2}c^{2}}.  \label{30}
\end{equation}%
Therefore, the solution of the problem of stationary states in RQM is the
ordinary De-Broglie wave. In the next section we develop this approach in
application to a quantum bosonic string.

\section{\textbf{QUANTUM ACTION PRINCIPLE IN STRING THEORY}}

We consider here a closed bosonic string parameterized by two parameters $%
\left( \tau ,\sigma \right) \in \left[ 0,1\right] \times \left[ 0,\pi \right]
$ with classical action in the canonical form \cite{GSW}:
\begin{equation}
I=\int\limits_{0}^{1}d\tau \int\limits_{0}^{\pi }d\sigma \left[ p_{\mu }%
\overset{\cdot }{x}^{\mu }-N_{1}H_{1}-N_{2}H_{2}\right] ,  \label{31}
\end{equation}

\begin{equation}
H_{1}\equiv \left( p_{\mu }+\gamma x_{\mu }^{\prime }\right)
^{2},H_{1}\equiv \left( p_{\mu }-\gamma x_{\mu }^{\prime }\right) ^{2},
\label{32}
\end{equation}%
where the dot denotes the derivative with respect to $\tau $, and the stroke
denotes the derivative with respect to $\sigma $. The constraints (\ref{32})
are in involution, i,e., their Poisson brackets (PB) equal zero. This is the
consequence of covariance of the theory with respect to arbitrary
transformations of coordinates $\left( \tau ,\sigma \right) $ on a world
surface of a string. In order to guarantee the classical character of the
dynamics of the $x^{0}\left( \sigma \right) $-coordinates of a string points
at the quantum level, we enlarge the set of variational parameters in the
classical action as follows:
\begin{eqnarray}
I &=&\int\limits_{0}^{1}d\tau \int\limits_{0}^{\pi }d\sigma \left\{ p_{\mu
}\overset{\cdot }{x}^{\mu }\right.  \label{33} \\
&&-N_{1}\left[ d_{1}^{2}-\left( p_{i}+\gamma x_{i}^{\prime }\right) ^{2}%
\right] -N_{2}\left[ d_{2}^{2}-\left( p_{i}-\gamma x_{i}^{\prime }\right)
^{2}\right]  \notag \\
&&\left. -\lambda _{1}\left[ d_{1}-\left( p_{0}+\gamma x_{0}^{\prime
}\right) \right] -\lambda _{2}\left[ d_{2}-\left( p_{0}-\gamma x_{0}^{\prime
}\right) \right] \right\} .  \notag
\end{eqnarray}%
Notice that the expressions in the square brackets under the integral are in
involution because $d_{1,2}$ have zero PB with any dynamical variable.
Therefore, the covariance of the modified theory is not broken. This will
help us to simplify the subsequent consideration, assuming that $N_{1,2}$
are independent on $\tau $. Corresponding Schr\"{o}dinger equation has the form
\begin{eqnarray}
i\hbar \frac{\partial \psi }{\partial \tau } &=&\int\limits_{0}^{\pi
}d\sigma \left[ N_{1}d_{1}^{2}+N_{2}d_{2}^{2}+\lambda _{1}d_{1}+\lambda
_{2}d_{2}\right.  \label{34} \\
&&-\lambda _{1}\left( \frac{\hbar }{i}\frac{\delta }{\delta x^{0}}+\gamma
x_{0}^{\prime }\right) -\lambda _{2}\left( \frac{\hbar }{i}\frac{\delta }{%
\delta x^{0}}-\gamma x_{0}^{\prime }\right)  \notag \\
&&\left. -N_{1}\left( \frac{\hbar }{i}\frac{\delta }{\delta x^{i}}+\gamma
x_{i}^{\prime }\right) ^{2}-N_{2}\left( \frac{\hbar }{i}\frac{\delta }{%
\delta x^{i}}-\gamma x_{i}^{\prime }\right) ^{2}\right] \psi .  \notag
\end{eqnarray}

We separate quantum dynamics of the $x^{0}$-coordinates and spatial
coordinates of a string looking for a solution of the equation (\ref{34}) as
the product $\psi \left[ \tau ,x^{\mu }\left( \sigma \right) \right] =\psi
_{0}\left[ \tau ,x^{0}\left( \sigma \right) \right] \psi _{1}\left[ \tau
,x^{i}\left( \sigma \right) \right] $. Quantum dynamics of the $x^{0}$%
-coordinates is described by an equation,
\begin{eqnarray}
i\hbar \frac{\partial \psi _{0}}{\partial \tau } &=&\int\limits_{0}^{\pi
}d\sigma \left[ N_{1}d_{1}^{2}+N_{2}d_{2}^{2}+\lambda _{1}d_{1}+\lambda
_{2}d_{2}\right.  \label{35} \\
&&\left. -\lambda _{1}\left( \frac{\hbar }{i}\frac{\delta }{\delta x^{0}}%
+\gamma x_{0}^{\prime }\right) -\lambda _{2}\left( \frac{\hbar }{i}\frac{%
\delta }{\delta x^{0}}-\gamma x_{0}^{\prime }\right) \right] \psi _{0}.
\notag
\end{eqnarray}%
We look for its solution in the exponential form
\begin{equation}
\psi _{0}\left[ \tau ,x^{0}\left( \sigma \right) \right] =\exp \left\{ \frac{%
i}{\hbar }\chi \left[ \tau ,x^{0}\left( \sigma \right) \right] \right\} ,
\label{36}
\end{equation}
\begin{eqnarray}
\chi \left[ \tau ,x^{0}\left( \sigma \right) \right] &=&\chi _{0}\left( \tau
\right) +\int\limits_{0}^{\pi }d\sigma \chi _{1}\left( \tau ,\sigma \right)
x^{0}\left( \sigma \right)  \label{37} \\
&&+\frac{1}{2}\int\limits_{0}^{\pi }d\sigma \int\limits_{0}^{\pi }d%
\widetilde{\sigma }\chi _{2}\left( \tau ,\sigma ,\widetilde{\sigma }\right)
x^{0}\left( \sigma \right) x^{0}\left( \widetilde{\sigma }\right)  \notag
\end{eqnarray}%
with symmetric third coefficient: $\chi _{2}\left( \tau ,\sigma ,\widetilde{%
\sigma }\right) =\chi _{2}\left( \tau ,\widetilde{\sigma },\sigma \right) $,
and take at the initial moment $\tau =0$:
\begin{equation}
\chi \left[ \tau ,x^{0}\left( \sigma \right) \right] =\frac{i\hbar }{%
4\varepsilon ^{2}}\left( \int\limits_{0}^{\pi }d\sigma \left( x^{0}\right)
^{2}\right) ^{2}+\int\limits_{0}^{\pi }d\sigma p_{0}x^{0}.  \label{38}
\end{equation}%
It means, that we take the initial value of $x^{0}\left( \sigma \right) $
from the $\varepsilon $-neighbourhood of zero, and corresponding initial
momentum equals $p_{0}\left( \sigma \right) $. Substituting (\ref{36}) in
the equation (\ref{35}) and comparing coefficients in front of $x^{0}$%
-degrees in both sides, we obtain:
\begin{equation}
\chi _{2}\left( \tau ,\sigma ,\widetilde{\sigma }\right) =\chi _{2}\left(
0,\sigma ,\widetilde{\sigma }\right) =\frac{i\hbar }{2\varepsilon ^{2}}%
\delta \left( \sigma -\widetilde{\sigma }\right) ,  \label{39}
\end{equation}
\begin{eqnarray}
\chi _{1}\left( \tau ,\sigma \right) &=&-\gamma \int\limits_{0}^{\tau }d%
\widetilde{\tau }\left[ \lambda _{1}^{\prime }\left( \widetilde{\tau }%
\right) -\lambda _{2}^{\prime }\left( \widetilde{\tau }\right) \right] +p_{0}
\notag \\
&&+\frac{i\hbar }{2\varepsilon ^{2}}\int\limits_{0}^{\tau }d\widetilde{\tau
}\left[ \lambda _{1}\left( \widetilde{\tau }\right) +\lambda _{2}\left(
\widetilde{\tau }\right) \right] ,  \label{40}
\end{eqnarray}
\begin{eqnarray}
&&\chi _{0}\left( \tau \right)  \label{41} \\
&=&-\int\limits_{0}^{\tau }d\widetilde{\tau }\int\limits_{0}^{\pi }d\sigma
\left( N_{1}d_{1}^{2}+N_{2}d_{2}^{2}+\lambda _{1}d_{1}+\lambda
_{2}d_{2}\right)  \notag \\
&&-\int\limits_{0}^{\tau }d\widetilde{\tau }\int\limits_{0}^{\pi }d\sigma
\left( \lambda _{1}+\lambda _{2}\right) p_{0}  \notag \\
&&-\gamma \int\limits_{0}^{\pi }d\sigma \int\limits_{0}^{\tau }d\widetilde{%
\tau }\left( \lambda _{1}+\lambda _{2}\right) \int\limits_{0}^{\widetilde{%
\tau }}d\widetilde{\widetilde{\tau }}\left( \lambda _{1}^{\prime }-\lambda
_{2}^{\prime }\right)  \notag \\
&&+\frac{i\hbar }{2\varepsilon ^{2}}\int\limits_{0}^{\pi }d\sigma
\int\limits_{0}^{\tau }d\widetilde{\tau }\left( \lambda _{1}+\lambda
_{2}\right) \int\limits_{0}^{\widetilde{\tau }}d\widetilde{\widetilde{\tau }%
}\left( \lambda _{1}+\lambda _{2}\right) .  \notag
\end{eqnarray}%
We are ready to write the solution of the equation (\ref{35}) on the
interval $\tau \in \left[ 0,1\right] $ in terms of real functionals $\Lambda
_{x^{0}},R_{x^{0}}$ defined by the exponential representation (\ref{5}):
\begin{eqnarray}
&&\Lambda _{x^{0}}  \label{42} \\
&=&-\int\limits_{0}^{1}d\widetilde{\tau }\int\limits_{0}^{\pi }d\sigma
\left( N_{1}d_{1}^{2}+N_{2}d_{2}^{2}+\lambda _{1}d_{1}+\lambda
_{2}d_{2}\right)  \notag \\
&&+\int\limits_{0}^{1}d\widetilde{\tau }\int\limits_{0}^{\pi }d\sigma
\left( \lambda _{1}+\lambda _{2}\right) p_{0}+\int\limits_{0}^{\pi }d\sigma
p_{0}\widetilde{x}^{0}  \notag \\
&&-\gamma \int\limits_{0}^{\pi }d\sigma \int\limits_{0}^{1}d\widetilde{%
\tau }\left( \lambda _{1}+\lambda _{2}\right) \int\limits_{0}^{\widetilde{%
\tau }}d\widetilde{\widetilde{\tau }}\left( \lambda _{1}^{\prime }-\lambda
_{2}^{\prime }\right)  \notag \\
&&-\gamma \int\limits_{0}^{1}d\widetilde{\tau }\int\limits_{0}^{\pi
}d\sigma \left( \lambda _{1}^{\prime }-\lambda _{2}^{\prime }\right)
\widetilde{x}^{0},  \notag
\end{eqnarray}
\begin{equation}
R_{x^{0}}=-\frac{1}{4\varepsilon ^{2}}\int\limits_{0}^{\pi }d\sigma \left[
\widetilde{x}^{0}+\int\limits_{0}^{1}d\widetilde{\tau }\left( \lambda
_{1}+\lambda _{2}\right) \right] ^{2},  \label{43}
\end{equation}%
where $\widetilde{x}^{0}\left( \sigma \right) $ is a final distribution of
the $x^{0}$-coordinate along a string. $\Lambda _{x^{0}}$ is a part of a
quantum action corresponding to the dynamics of the $x^{0}$-coordinate of a
string. Conditions of the stationarity of $\Lambda _{x^{0}}$ with respect to
additional variables $d_{1,2}$ and $\lambda _{1,2}$ are the following:
\begin{equation}
2N_{1,2}d_{1,2}+\lambda _{1,2}=0,  \label{44}
\end{equation}
\begin{eqnarray}
0 &=&-d_{1}+p_{0}-\gamma \int\limits_{0}^{\tau }d\widetilde{\tau }\left(
\lambda _{1}^{\prime }-\lambda _{2}^{\prime }\right)  \label{45} \\
&&+\gamma \int\limits_{\tau }^{1}d\widetilde{\tau }\left( \lambda
_{1}^{\prime }+\lambda _{2}^{\prime }\right) +\left( \widetilde{x}%
^{0}\right) ^{\prime },  \notag \\
0 &=&-d_{1}+p_{0}-\gamma \int\limits_{0}^{\tau }d\widetilde{\tau }\left(
\lambda _{1}^{\prime }-\lambda _{2}^{\prime }\right)  \notag \\
&&-\gamma \int\limits_{\tau }^{1}d\widetilde{\tau }\left( \lambda
_{1}^{\prime }+\lambda _{2}^{\prime }\right) -\left( \widetilde{x}%
^{0}\right) ^{\prime }  \notag
\end{eqnarray}%
Taking derivatives of the equations (\ref{45}) with respect to $\tau $
and taking into account (\ref{44}), we obtain
\begin{equation}
\frac{\partial }{\partial \tau }\left( \frac{\lambda _{1}}{2N_{1}}\right)
=-2\gamma \lambda _{1}^{\prime },\frac{\partial }{\partial \tau }\left(
\frac{\lambda _{2}}{2N_{2}}\right) =2\gamma \lambda _{2}^{\prime }.
\label{46}
\end{equation}%
Taking $\tau =0$ in the equations (\ref{45}), we obtain:
\begin{equation}
\lambda _{1,2}\left( 0,\sigma \right) =\mp 2N_{1,2}\left( 0,\sigma \right)
p_{0}\left( \sigma \right) .  \label{47}
\end{equation}%
Therefore, we have two wave equations (\ref{46}) for $\lambda _{1,2}$ - a
wave to the right, and a wave to the left, and corresponding initial data (%
\ref{47}) for both waves.

From the other hand, according to (\ref{43}) in the limit $\varepsilon
\rightarrow 0$, we have
\begin{equation}
\widetilde{x}^{0}\left( \sigma \right) =-\int\limits_{0}^{1}d\widetilde{%
\tau }\left( \lambda _{1}\left( \tau ,\sigma \right) +\lambda _{2}\left(
\tau ,\sigma \right) \right) .  \label{48}
\end{equation}%
The set of the equations (\ref{46}),(\ref{47}), and (\ref{48}) defines $%
p_{0}\left( \sigma \right) $ as a functional of $N_{1,2}\left( \tau ,\tau
\right) $, and $\widetilde{x}^{0}\left( \sigma \right) $. This functional is
homogeneous of the $+1$ degree on $x^{0}\left( \sigma \right) $, and of the $%
-1$ degree on $N_{1,2}\left( \tau ,\sigma \right) $. This result is an
analog of the equation (\ref{28}) in the case of a relativistic particle.
Collecting all the results together, we obtain the stationary value of $%
\Lambda _{x^{0}}$ (with respect to the additional variables $d_{1,2}$ and $\lambda
_{1,2}$) as a functional, homogeneous of the $+2$ degree on $x^{0}\left(
\sigma \right) $, and of the $-1$ degree on $N_{1,2}\left( \tau ,\sigma
\right) $.

Let us turn to the remained second multiplier of a wave function, which
describes the dynamics of spatial coordinates $x^{i}$ of a string, and obeys
the Schr\"{o}dinger equation
\begin{eqnarray}
&&i\hbar \frac{\partial \psi _{1}}{\partial \tau }  \label{49} \\
&=&-\left[ N_{1}\left( \frac{\hbar }{i}\frac{\delta }{\delta x^{i}}+\gamma
x_{i}^{\prime }\right) ^{2}+N_{2}\left( \frac{\hbar }{i}\frac{\delta }{%
\delta x^{i}}-\gamma x_{i}^{\prime }\right) ^{2}\right] \psi _{1}.  \notag
\end{eqnarray}%
Subsequent results will depend on a concrete physical problem under consideration.
We are interested in stationary states of a quantum string. The
Hamiltonian in the right hand side of the equation (\ref{49}) describes a
system of bound oscillators. Therefore, the stationary states we are looking
for are excitations of these oscillators. However coefficients $N_{1,2}\left(
\sigma \right) $ in the Hamiltonian remain arbitrary up to now. For any
eigenfunction of the Hamiltonian $\left\vert n\right\rangle ,n=\left(
n_{1},n_{2},...\right) $ with the energy $E_{n}$, the transition amplitude $%
K_{nx^{i}}\equiv \left\langle out,n\right. \left\vert in,n\right\rangle $
equals to
\begin{equation}
K_{nx^{i}}=\exp \left( -\frac{i}{\hbar }\int\limits_{0}^{1}d\widetilde{\tau
}E_{n}\right) .  \label{50}
\end{equation}%
Notice, that the eigenvalues $E_{n}$ of the Hamiltonian, as the Hamiltonian
itself, are homogeneous functions of the first degree on $N_{1,2}$.
Therefore, the same degree of homogeneity has the spatial part of a quantum
action for stationary states:
\begin{equation}
\Lambda _{x^{i}}=-\int\limits_{0}^{1}d\widetilde{\tau }E_{n}.  \label{51}
\end{equation}%
This result is analogous to the Eq.(\ref{23}) in the case of a relativistic
particle. Collecting both parts of the quantum action, we write it as the
sum:
\begin{equation}
\Lambda =\Lambda _{x^{0}}+\Lambda _{x^{i}}.  \label{52}
\end{equation}%
This result is analogous to the Eq.(\ref{29}) in the case of a relativistic
particle. We get a quantum analog of the Lagrangian form of the classical
action of a bosonic string.

Now, we are ready to complete the formulation of QAP by solving the
conditions of the stationarity (\ref{6}) with respect to remained arbitrary
Lagrangian multipliers $N_{1,2}$. The structure of the action (\ref{52}) is
such, that stationary values of $N_{1,2}$ will be homogeneous functionals
of first degree on the $\widetilde{x}^{0}$. Consequently, the stationary
value of the action (\ref{52}) itself will be a homogeneous functional of
the first degree on the $\widetilde{x}^{0}$. At last, we can take the
boundary distribution $\widetilde{x}^{0}\left( \sigma \right) =const$ along a
string. Then, the stationary value of the action (\ref{52}) will be
proportional $\widetilde{x}^{0}$:
\begin{equation}
\Lambda =\widetilde{x}^{0}W_{n},  \label{53}
\end{equation}%
where $W_{n}$ is the energy of string in the stationary state $\left\vert
n\right\rangle $. It is obvious, that the stationary value of the energy $%
W_{n}$ is a homogeneous function of the $1/2$-degree on quantum numbers $n$.
Therefore, we have an analog of the square route in (\ref{30}).

Is the stationary state $\left\vert n\right\rangle $ in fact stationary?
Yes, it is. For arbitrary values $N_{1,2}\left( \sigma \right) $ the
transition amplitude $\left\langle out,m\right. \left\vert in,n\right\rangle
=0$, if $m\neq n$. This zero value will be conserved in the stationary point
with respect to $N_{1,2}\left( \sigma \right) $ (if it exists). Therefore,
the probabilistic interpretation of the free string quantum dynamics in our
framework is trivial, namely, a stationary state $\left\vert n\right\rangle $ remains
unchanged with the unit probability. Notice, that the stationary states $%
\left\vert n\right\rangle $ are, in fact, non-orthogonal.

\section{\textbf{CONCLUSIONS}}

In the present work we propose to impose the conditions of the
stationarity of the classical action with respect to Lagrangian multipliers
on to the quantum action. These conditions of the stationarity, i.e., constraints,
are not to be solved at the classical level, or transformed in quantum
constraints, as in ordinary covariant quantum theory. We must form the
transition amplitude of a real quantum process and take its real phase as
a quantum action (of the process). The quantum action is an analog of the
Lagrangian form of the classical action. The conditions of the stationarity of the
quantum action with respect to the Lagrangian multipliers connect them (or more
precisely, their invariant combinations) with kinematical parameters of the
process.The only memory about quantum anomalies which arise in the ordinary
covariant quantum theory is the question: what is the precision to which
the Lagrangian multipliers are fixed in QAP? In any case, the transition amplitudes
of quantum processes and stationary states are well defined in the
Minkowsky space of arbitrary dimension.

We are thanks V. A. Franke and A. V. Goltsev for useful discussions.





\begin{thebibliography}{9}
\bibitem{D} P.A.M.Dirac, \textit{Lectures on Quantum Mechanics} (Paperback,
22-mar-01, Dover Publication).

\bibitem{BD} James D. Bjorken, Sidney D. Drell, \textit{Relativistic Quantum
Mechanics} (McGraw-Hill Book Company 1976).

\bibitem{GLL1} Natalya Gorobey, Alexander Lukyanenko, and Inna Lukyanenko,
arXiv: 1010.3824v1[quant-ph] (19 October 2010).

\bibitem{GLL2} Natalya Gorobey, and Alexander Lukyanenko, and Inna
Lukyanenko, arXiv: 1012.1719v1[quant-ph] (8 December 2010).

\bibitem{D1} P.Dirac, The Physical Rev., second series, v.74, 1948,
p.817-830.

\bibitem{GSW} Michael B. Green, John H. Schwarz, and Edward Witten, \textit{%
Superstring Theory} (Cambridge Univ. Press, N.Y., 1987).
\end{thebibliography}
\end{document}